\documentclass[aps,twocolumn,preprintnumbers,nofootinbib,prc,superscriptaddress]{revtex4-2}
\usepackage{graphicx,color}
\usepackage{float}
\usepackage{amsmath,amssymb,amsfonts}
\usepackage{braket}
\usepackage{bm}
\usepackage[pdftex,colorlinks=true, linkcolor = blue, citecolor=blue,urlcolor=blue, bookmarksnumbered=true, bookmarksopen=true]{hyperref}
\usepackage[normalem]{ulem}
\begin{document}

\title{Simplified projection on total spin zero for state preparation on quantum computers}

\author{E. Rule}
\affiliation{Department of Physics, University of California, Berkeley, CA 94720, USA}
\affiliation{Los Alamos National Laboratory, Theoretical Division, Los Alamos, New Mexico 87545, USA}
\author{I. Stetcu}
\affiliation{Los Alamos National Laboratory, Theoretical Division, Los Alamos, New Mexico 87545, USA}
\author{J. A. Carlson}
\affiliation{Los Alamos National Laboratory, Theoretical Division, Los Alamos, New Mexico 87545, USA}

\date{\today}
\preprint{LA-UR-24-20109 v2, N3AS-24-032}

\begin{abstract}

We introduce a simple algorithm for projecting on $J=0$ states of a many-body system by performing a series of rotations to remove states with angular momentum projections greater than zero. Existing methods rely on unitary evolution with the two-body operator $J^2$, which when expressed in the computational basis contains many complicated Pauli strings requiring Trotterization and leading to very deep quantum circuits. Our approach performs the necessary projections using the one-body operators $J_x$ and $J_z$. By leveraging the method of Cartan decomposition, the unitary transformations that perform the projection can be parameterized as a product of a small number of two-qubit rotations, with angles determined by an efficient classical optimization. Given the reduced complexity in terms of gates, this approach can be used to prepare approximate ground states of even-even nuclei by projecting onto the $J=0$ component of deformed Hartree-Fock states. We estimate the resource requirements in terms of the universal gate set $\{H,S,\textsc{cnot},T\}$ and briefly discuss a variant of the algorithm that projects onto $J=1/2$ states of a system with an odd number of fermions.

\end{abstract}
\maketitle

\section{Introduction}
Mean-field approaches to many-body calculations have a long and illustrious history that can continue even in an early quantum era, when quantum hardware will reach the point to offer advantage in generic quantum many-body systems over the classical methods in describing many-body dynamics, but will still not be able to handle relatively large coherence times required in applications to realistic problems in quantum chemistry or nuclear physics. Preparation of an initial quantum state remains an inherently difficult problem that involves potentially long propagation times and hence deep quantum circuits, despite progress made in devising algorithms for state preparation \cite{AbramsLloyd1997,Abrams1999,Poulin2009,Ge2018,Motta:2020,Choi2021,Dong2022,Jouzdani2022,Stetcu-2023proj}. 

In nuclear physics, ground states of many-body systems can be successfully approximated by mean-field solutions, especially if these solutions are projected on good quantum numbers, thus restoring symmetries of the Hamiltonian that are broken by the mean-field approximation \cite{Sheikh:2019qdz}. Including particle-hole correlations via unitary coupled-cluster-like methods can also improve the quality of the solutions \cite{BARTLETT1989133,https://doi.org/10.1002/qua.21198}, though there is no guarantee that the symmetries broken by the mean-field approach are fully restored. Symmetry restoration has been combined with variational quantum eigensolver methods \cite{Peruzzo:2013bzg,McClean:2015vup} to improve the determination of ground-state and excited-state energies of many-body systems \cite{PhysRevResearch.2.043142,PhysRevC.105.024324,PhysRevC.107.034310}. As symmetries play a major role in many-body dynamics, it is desirable to perform symmetry restoration during state-preparation algorithms before the prepared state is used as input to study such dynamics.

In this paper, we propose an algorithm that efficiently performs a projection to $J=0$ quantum numbers for a system with an even number of fermions.  As this approach is based on a projection algorithm \cite{Stetcu-2023proj}, restriction to $J=0$ is necessary; only in this case is the initial success probability for the state preparation --- given by the overlap of the projected state with the initial trial wave function --- preserved. In other words, the amplitude of the $J=0$ component is unchanged throughout the process, as states with $J>0$ are progressively removed. Given that the ground states of all even-even nuclei have total angular momentum $J=0$, the proposed method has applications to a large number of nuclei, both close to and far from stability. 

In Ref. \cite{Stetcu-2023proj}, a method for projecting on the spin quantum number was introduced in which one uses a series of correlated time evolutions followed by measurements of an ancilla qubit. Other methods based on quantum-phase estimation \cite{PhysRevLett.125.230502,PhysRevA.104.062435,10.1140/epja/s10050-022-00911-7} can be used to project a deformed initial state onto the complete eigenbasis of $J^2$ and $J_z$, whereas the algorithm proposed in Ref. \cite{Stetcu-2023proj} targets a specific $J$ component. In either case, the implementation of these methods on quantum hardware remains challenging as they typically require a large number of gates --- in particular \textsc{cnot} gates --- as a consequence of performing unitary evolution with the two-body operator $J^2$. For nuclear systems where both protons and neutrons are involved in the calculation, proton-neutron cross terms generate a large number of complicated Pauli strings, leading to deep quantum circuits. We will quantify this statement more precisely in Sec. \ref{sec:resources}. For now, it suffices to say that it would be desirable to use the one-body operators $J_x$ and $J_z$ to perform the necessary projections, as these operators act separately on protons and neutrons, allowing one to factorize the problem. 

This is exactly the approach we pursue in this work: Our proposed algorithm progressively removes states that transform non-trivially under rotations about the $x$ and $z$ axes, leaving behind only states invariant under all rotations: those with total angular momentum $J=0$. Using the method of Cartan decomposition \cite{KHANEJA200111,10.1063/1.2008210,Kokcu:2021ctj}, unitary evolution with the one-body operators $J_x$ and $J_z$ can be implemented very efficiently without Trotterization, significantly reducing the resource requirements compared to algorithms based on the two-body operator $J^2$. In light of this improvement --- which comes at the cost of more mid-circuit measurements --- the proposed algorithm could be used in practical calculations in the near future. A combination of this method and a variational approach would also most likely provide better approximations for ground states with considerably lower resource requirements.

Given our personal bias towards nuclear physics, the trial wave functions that we consider in our numerical examples are Hartree-Fock states obtained from nuclear shell-model interactions. Nonetheless, the projection algorithm that we present is more general and should be applicable to a wide range of many-body fermionic systems in nuclear physics, quantum chemistry, and condensed matter.

This paper is organized as follows: In Sec. \ref{sec:theory}, we present the proposed algorithm for projecting $J=0$ states from trial wave functions containing a mixture of total angular momentum eigenstates. We describe in detail how this can be achieved through a series of correlated rotations using the operators $J_x$ and $J_z$. We also briefly discuss a variation of the method that can be used to project to the $J=1/2$ component of a system with an odd number of fermions. In Sec. \ref{sec:cartan}, we demonstrate that Cartan decomposition can be used to efficiently encode the required rotation operators. In Sec. \ref{sec:circuit}, we provide an overview of the implementation of our algorithm on quantum hardware, including an example circuit. In Sec. \ref{sec:results}, we present several numerical examples from nuclear physics: Sec. \ref{sec:convergence} demonstrates the convergence of our proposed algorithm, and Sec. \ref{sec:resources} compares the estimated resource requirements to those of the algorithm introduced in Ref. \cite{Stetcu-2023proj}, which performs the projection with a two-body operator. We summarize our work in Sec. \ref{sec:summary}. 

To aid the reader, two Appendices are included: Appendix \ref{app:cartan} contains more details on the Cartan decomposition, and Appendix \ref{app:numerical_params} provides the reader with numerical values for the corresponding transformation coefficients.

\section{Theoretical approach}
\label{sec:theory}
A deformed trial wave function can always be written as a superposition of states with good total angular momentum $J$ and magnetic projection $M$,
\begin{equation}
    \ket{\psi_\mathrm{trial}}=\sum_{n,J,M} A_{nJM} \ket{n\;J\;M},
    \label{eq:trial}
\end{equation}
where $A_{n JM}$ is the amplitude of each basis state, and the index $n$ represents all additional quantum numbers. A rotation of the trial state by a set of Euler angles $\Omega=(\alpha,\beta,\gamma)$ produces a new (deformed) state \cite{Varshalovich:1989,Ring:2004}
\begin{widetext}
    
\begin{equation}
\exp\left(-iJ_z\alpha\right)\exp\left(-iJ_y\beta\right)\exp\left(-iJ_z\gamma\right)\ket{\psi_\mathrm{trial}} =
    \sum_{n,J,M} A_{n JM} \sum_{M'} {\cal D}^{J*}_{M'M} (\Omega) \ket{n\;J\;M} = \sum_{n,J,M'} A'_{n JM'} \ket{n\;J \;M'},
    \label{eq:transform}
\end{equation}
\end{widetext}
where ${\cal D}^{J*}_{M'M} (\Omega)$ are the Wigner D-functions, and $J_y$ and $J_z$ are the operators corresponding to spin projection on $y$ and $z$ axes, respectively. Thus, a rotation of the trial wave function changes the relative weight of states with the same total angular momentum $J$ but different angular momentum projections $M$ via a unitary transformation; that is,
\begin{equation}
    A'_{nJM'}\equiv\sum_M {\cal D}^{J*}_{M'M} (\Omega) A_{n JM}.
\end{equation} 
We make the following observations: 
\begin{itemize}
    \item[\textit{(i)}] The coefficients of $J=0$ states are invariant under all rotations, as ${\cal D}^{0}_{00}(\Omega)=1$.
    \item[\textit{(ii)}] If we remove states with $M\neq 0$ from the trial wave function, then subsequent rotations will generically cause states with $(J>0,M\neq0)$ to acquire a non-zero amplitude, while simultaneously suppressing the amplitude of $(J>0,M=0)$ states, given the unitary nature of the transformation.
\end{itemize}
The second point is crucial to our algorithm because we cannot directly remove states with $(J> 0,M=0)$ without also eliminating the $(J=0,M=0)$ component of interest. However, we can rely on the fact that rotations distribute the amplitude of the $(J> 0,M=0)$ component into the corresponding $(J> 0,M\neq 0)$ components, which can be readily eliminated, as we will now describe.

\subsection{Projection with $J_z$ and $J_x$}
\label{sec:projection}
\begin{figure*}
    \centering
    \includegraphics[scale=0.84]{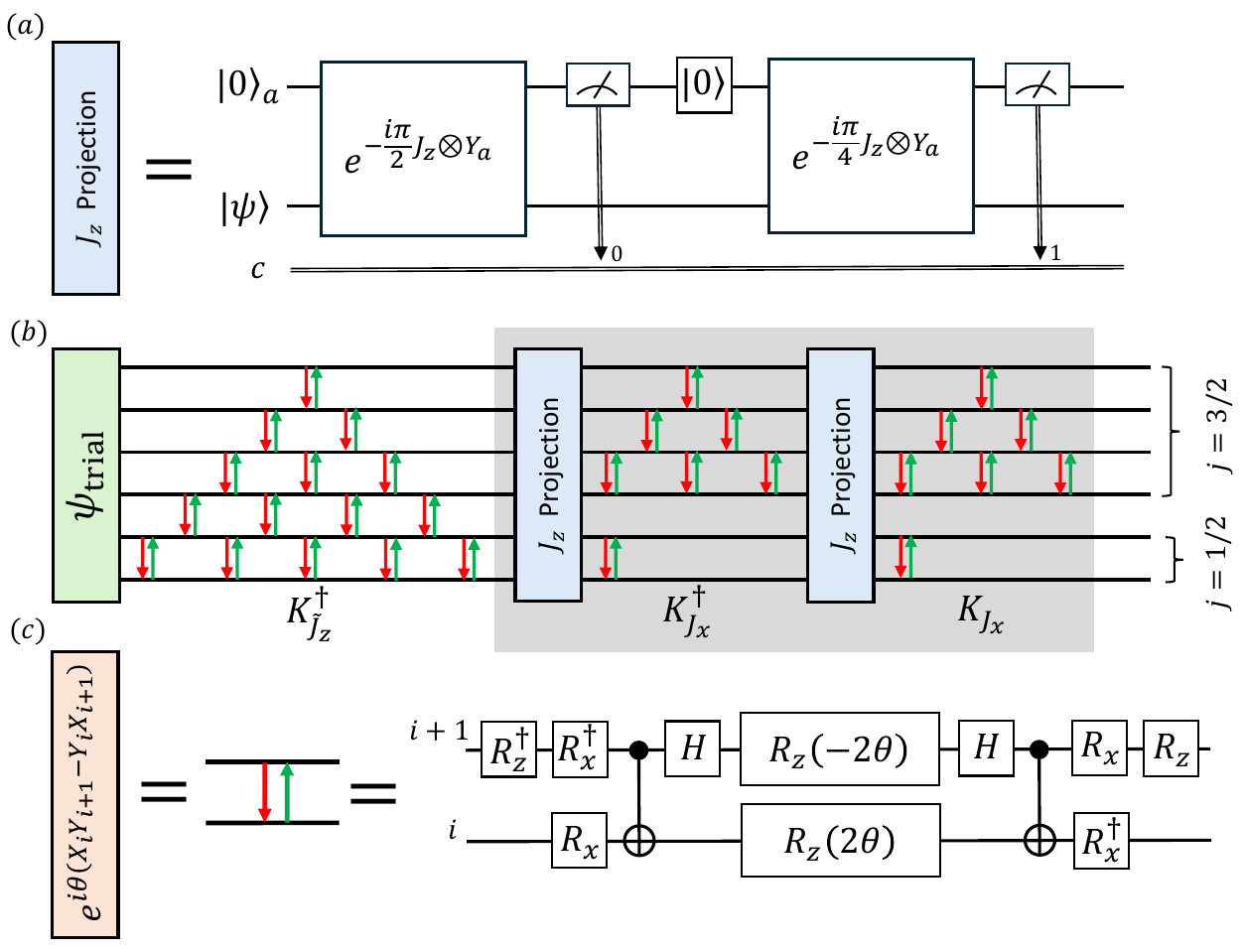}
    \caption{Example quantum circuit that implements the proposed projection algorithm: (a) Sub-circuit that performs 2 projections, requiring a single ancilla qubit in total and one mid-circuit measurement/reset for each projection. (b) A trial wave function is prepared in a deformed basis. The first unitary transformation $K^\dag_{\tilde{J}_z}$ transforms to the conventional spherical basis where $J_z$ is diagonal. Each pair of red and green arrows represents the operator $\exp[i\theta (X_iY_{i+1}-Y_iX_{i+1})]$. After this transformation is performed, the top 4 (bottom 2) qubits correspond to single-particle states with good angular momentum quantum numbers $j=3/2$ ($j=1/2)$. Next, states with total magnetic projection $M\neq 0$ are progressively removed by time-evolving with $J_z$ for different periods $t=\frac{\pi}{2},\frac{\pi}{4},\ldots$\;. Subsequent transformations to diagonalize $J_x$ take advantage of the fact that single-particle states with different total $j$ do not mix; in this example, $K_{J_x}=K_{J_x}^{(1/2)}\otimes K_{J_x}^{(3/2)}$. The shaded grey region denotes the circuit that defines a single iteration of the projection algorithm. (c) Sub-circuit that implements the two-qubit operator $\exp[i\theta (X_iY_{i+1}-Y_iX_{i+1})]$. The $R_x$ and $R_z$ gates represent $R_x(\pi/2)$ and $R_z(\pi/2)$, respectively.}
    \label{fig:circuit_sketch}
\end{figure*}

The algorithm we propose is similar to the one introduced in Ref. \cite{Stetcu-2023proj}, but instead of using $J^2$ to project on good angular momentum, we do so with the rotation operators $J_z$ and $J_x$. Let's first consider an even-even nucleus so that the total angular momentum assumes integer values. Using a single ancilla qubit $a$ initialized in the $\ket{0}$ state, we perform a series of unitary transformations $\exp\left(-it_i J_z\otimes Y_a\right)$ with different times $t_i$, as illustrated in Fig. \ref{fig:circuit_sketch} (a). After each unitary evolution, the state of the system becomes
\begin{equation}
\begin{split}
\ket{\psi(t_i)}&=\exp\left(-it_i J_z\otimes Y_a\right)\ket{\psi}\otimes \ket{0}_a\\
&=\cos(J_zt_i)\ket{\psi}\otimes\ket{0}_a+\sin(J_zt_i)\ket{\psi}\otimes\ket{1}_a,
\end{split}
\label{eq:unitary_proj_operator}
\end{equation}
which we can project into a particular subspace by performing a mid-circuit measurement on the ancilla qubit.\footnote{In lieu of mid-circuit measurements/resets of a single ancilla, one can defer all measurements until the end of the circuit by introducing a new ancilla qubit for each application of Eq. \eqref{eq:unitary_proj_operator}.} Specifically, if we evolve for time $t_1=\pi/2$ and then measure the ancilla qubit to be in state $\ket{0}_a$, we have eliminated from the new state all components with $M=\pm 1,\pm 3,\pm 5,\ldots$\;. The non-zero even-$M$ components of $\ket{\psi}$ cannot all be removed in a single step; for example, if we repeat the projection procedure with time $t_2=\pi/4$, we eliminate states with $M=\pm 2,\pm 6,\pm 10,\ldots$\;. Subsequent projections with time $t_i=\pi/(2^i)$ will remove states with $M=(2m+1)2^{i-1}$ for all $m\in \mathbb{Z}$.

Depending on the number of particles and the quantum numbers of the available single-particle states, there is a maximal value  of $|M|$ that can theoretically be present in $\ket{\psi}$.  However, as we demonstrate numerically in Sec. \ref{sec:results}, not all allowed values of $M\neq 0$ need to be eliminated at once in order to successfully project to the $J=0$ component. In fact, even if we were to remove all $M\neq 0$ states from $\ket{\psi}$, the result would be a state with quantum numbers $(J,M=0)$ for all allowed values of $J$. To overcome this, we must rely on observation $(ii)$ above and perform a rotation that will mix ($J> 0,M=0)$ states into ($J> 0,M\neq 0)$ states, thereby reducing the weight of the unwanted $M=0$ components. Depending on the rotation, this transformation can also dilute the relative weights of certain $M \neq 0$ components that were not explicitly removed via the initial projections. As the target $(J=0,M=0)$ component is invariant under all rotations, its amplitude remains unchanged throughout this process.

Therefore, after performing the desired number of projections with $J_z$, we next apply a series of unitary transformations $\exp\left(-it_i J_x\otimes Y_a\right)$ that allows us to eliminate particular values of angular momentum projections along the $x$ axis while simultaneously diluting the weights of $J\neq 0$ states as described above.\footnote{Equivalently, one could employ the operator $J_y$. We choose to use $J_x$ as its matrix elements are purely real in our conventions. See discussion in App. \ref{app:cartan}.} The elimination procedure for $J_x$ is exactly analogous to that described above for $J_z$; for example, evolving for time $t_1=\pi/2$ eliminates from $\ket{\psi}$ all odd values of the magnetic projection along the $x$ axis. As before, it is not necessary to remove all theoretically allowed values of the magnetic projection. This procedure is then repeated, projecting again using $J_z$ followed by $J_x$. With each successive iteration, more unwanted components are removed from the trial wave function until only the $J=0$ component remains.

Similar rationale can be used to devise an algorithm for projecting onto the $J=1/2$ component of a system with an odd number of fermions. While in this case the relevant Wigner D-matrix $\mathcal{D}^{1/2}_{M'M}(\Omega)$ is non-trivial, keeping only states with $M=\pm1/2$ will eventually filter out all components with $J>1/2$. The required evolution times are $t=\pi/(2\tilde{M})$, with $|\tilde{M}|>1/2$ leading to the removal of states with $M=(2m+1)\tilde{M}$, for all $m\in \mathbb{Z}$. After each evolution, the amplitude of the $M=\pm1/2$ components will be reduced by $\cos(\pi/(4\tilde{M}))$, but a renormalization will occur after the measurement (just like in the $J=0$ projection), so that overall the success probability will be preserved in the absence of other errors. It can be easily checked that the renormalization factor $\cos(\pi/(4\tilde{M}))$ is the highest of the remaining $M$ components, which are each rescaled by $\cos(M\pi/(2\tilde{M}))$. Therefore, no special order needs to be followed when removing $|M|> 1/2$ states, just like in the case of $J=0$ states. 

It is similarly true that not all $|M|>1/2$ states need to be removed at each step of the algorithm. Again, we rely on the fact that alternating rotations by $J_z$ and $J_x$ will dilute the relative strength of the $(J>1/2, |M|=1/2)$ components that cannot be directly removed. After sufficiently many iterations have been performed, only the $(J=1/2,M=\pm 1/2)$ states will remain. In practice, one usually seeks to project onto one particular magnetic component, either $M=+1/2$ or $M=-1/2$, of the $J=1/2$ multiplet. To achieve this, one can perform a final $J_z$ rotation with a non-trivial phase factor $\exp\left[-i(t_i J_z+\delta_i)\otimes Y_a\right]$ where $t_i = \pi/2$ and $\delta_i=\mp \pi/4$ to project onto either $M=\pm 1/2$.

\subsection{Cartan decomposition}
\label{sec:cartan}
The algorithm described in the previous section requires one to act on the trial wave function with rotation operators $J_x$ and $J_z$ for the many-body fermion system. If one works in the spherical basis and applies the standard Jordan-Wigner mapping, then spin up/down of the qubits encodes the occupation of fermionic single-particle states with good angular momentum $\ket{j\;m_z}$. In such case, the operator $J_z$ is diagonal in the computational basis and is straightforward to implement on quantum hardware. 

In practice, however, the trial wave function may be very complicated to encode in the spherical basis while being trivial to encode in the deformed basis where it was derived. The trade off is that $J_z$ must then be expressed in the deformed basis and in general will no longer be diagonal. We denote the non-diagonal form of the operator as $\tilde{J}_z$ to indicate that it is expressed in the deformed basis. This is the scenario we consider in our numerical examples below: Our trial wave functions are Hartree-Fock states expressed in terms of a deformed shell-model basis. We choose to prepare the trial wave function in the deformed basis, where $\tilde{J}_z$ is not diagonal. Therefore, the first step of our algorithm is to perform an efficient transformation from the deformed to the spherical basis, thus diagonalizing $\tilde{J}_z$.\footnote{If one begins with a trial wave function expressed in the spherical basis, this first step can be omitted. The remainder of our algorithm will apply just the same.} 

Of course, there must exist some unitary transformation $K_{\tilde{J}_z}$ that satisfies 
\begin{equation}
    \tilde{J}_z=K_{\tilde{J}_z}J_zK^\dag_{\tilde{J}_z},
    \label{eq:unitary_decomp}
\end{equation}
but a priori we have no notion of how efficiently $K_{\tilde{J}_z}$ can be implemented on a quantum computer. Applying the method of Cartan decomposition \cite{Kokcu:2021ctj}, $K_{\tilde{J}_z}$ can be expressed as a product of unitary two-qubit operations,
\begin{equation}
    K(\vec{\theta})=\prod_{\ell=1}^{n-1}\prod_{i=\ell}^1 e^{i\theta_{i,\ell}\left(X_iY_{i+1}-Y_iX_{i+1}\right)},
    \label{eq:K_op_basis}
\end{equation}
where the values of the $\theta_{i,\ell}$ parameters are determined by solving a classical optimization problem. For a system of $n$ qubits, the transformation $K_{\tilde{J}_z}$ is specified by $n(n-1)/2$ parameters. The necessary optimization requires only a local extremum of the relevant cost function. Both the cost function and its gradient can be efficiently evaluated, allowing one to employ standard optimization routines, such as the Broyden-Fletcher-Goldfarb-Shanno (BFGS) algorithm. Further details are given in Appendix \ref{app:cartan}. 

As the rotation operators conserve proton/neutron number, we can write
\begin{equation}
    \tilde{J}_z=\tilde{J}_{z,p}\oplus\tilde{J}_{z,n},
\end{equation}
and separately perform the decomposition in Eq. \eqref{eq:unitary_decomp} for each particle species, thereby reducing the complexity of each unitary transformation and decreasing the computational cost of the required classical optimizations. Separate transformations can then be applied to the proton and neutron components of the trial wave function,
\begin{equation}
    K_{\tilde{J}_z}\ket{\psi_\mathrm{trial}}=K_{\tilde{J}_{z,p}}\ket{\psi_\mathrm{trial,p}}\otimes K_{\tilde{J}_{z,n}}\ket{\psi_\mathrm{trial,n}}.
\end{equation}
The same approach can also be used to transform $J_x$, whose diagonal component must be identical to $J_z$,
\begin{equation}
    J_x=K_{J_x}J_zK^\dag_{J_x}.
    \label{eq:unitary_decomp_Jx}
\end{equation}
Like $K_{\tilde{J}_z}$, the unitary transformation $K_{J_x}$ can be expressed as a product of two-qubit operations of the form in Eq. \eqref{eq:K_op_basis}. After splitting the operator into proton and neutron components (labeled by $N=n,p$), we can further simplify the transformation $K_{J_{x,N}}$ using the fact that $J_{x,N}$ is block diagonal with respect to single-particle angular momentum $j$. That is, $J_{x,N}$ does not connect single-particle states $\ket{j\;m}$ with different $j$. Letting $P$ denote the set of single-particle $j$ values corresponding to the fermionic system of interest, we may write 
\begin{equation}
    J_{x,N}=\bigoplus_{j\in P}J_{x,N}^{(j)},
\end{equation}
where $J_{x,N}^{(j)}$ is the angular momentum operator restricted to the single-particle subspace with total angular momentum $j$. As a result, the unitary transformation that diagonalizes $J_{x,N}$ can be factorized as
\begin{equation}
    K_{J_{x,N}}=\bigotimes_{j\in P}\,K_{J_{x,N}}^{(j)}.
\end{equation}
As discussed below in Sec. \ref{sec:resources}, this factorization significantly reduces the depth of the corresponding quantum circuit (and reduces the computational cost of the requisite classical optimizations). An equivalent factorization of $K_{\tilde{J}_z}$ is generally not possible because $\tilde{J}_z$ is not block diagonal in the deformed basis.

The precise values of the parameters $\theta_{i,\ell}$ in Eq. \eqref{eq:K_op_basis} that determine $K_{\tilde{J}_z}$ depend on the trial wave function under consideration, as they encode the transformation between the deformed and spherical bases. On the other hand, the transformation $K^{(j)}_{J_x}$ is defined with respect to the usual spherical basis, and therefore the same set of $\theta_{i,\ell}$ parameters can be used for different trial wave functions. One possible set of numerical coefficients specifying the transformation $K^{(j)}_{J_x}$ for several values of single-particle $j$ is given in Table \ref{tab:KJx_params} of Appendix \ref{app:numerical_params}.

\subsection{Quantum Circuit}
\label{sec:circuit}
Combining the transformations $K_{\tilde{J}_z}$ and $K_{J_x}$ with the projection method described in Sec. \ref{sec:projection}, we can now describe our algorithm in full. An example quantum circuit is illustrated in Fig. \ref{fig:circuit_sketch} for a fermionic system consisting of 6 total single-particle states carrying angular momentum $j=1/2$ (2 states) and $j=3/2$ (4 states). The basic procedure is the following:
\begin{enumerate}
    \item Beginning from a trial wave function expressed in the deformed basis, we apply the unitary operator $K^{\dag}_{\tilde{J}_z}$, which transforms from the deformed to the spherical basis, where $J_z$ assumes its canonical diagonal form.
    \item We apply the projection algorithm for $J_z$ described in Sec. \ref{sec:projection}, evolving with $\exp(-it_iJ_z\otimes Y_a)$ for $t_i=\pi/2,\pi/4,...,\pi/(2^i)$ depending on the desired number of measurements per iteration.
    \item From the spherical basis where $J_z$ is diagonal, we apply the unitary operator $K_{J_x}^{\dag}$ that transforms to the corresponding spherical basis where $J_x$ is diagonal. 
    \item We apply the projection algorithm for $J_x$. Since we have transformed to the basis where $J_x$ is diagonal, the requisite quantum circuit is the same as for $J_z$ [shown in Fig. \ref{fig:circuit_sketch} (a)].
    \item We apply the transformation $K_{J_x}$, which returns the system to the spherical basis where $J_z$ is diagonal.
    \item We repeat steps (2--5) for the desired number of iterations.
\end{enumerate}
Thus, we define a single ``iteration'' of the algorithm as consisting of $N$ removals of non-zero projections on the $z$ axis followed by $N$ removals of non-zero projections on the $x$-axis, for a total of $N_\mathrm{proj}=2N$ measurements/removals per each iteration. Once the final iteration is complete, we do not need to apply the transformation $K_{\tilde{J}_z}$, which would take us back to the deformed basis. By omitting this term, we end the final iteration of the projection algorithm in the spherical basis, which is likely more convenient for additional manipulations of the prepared state. Of course, if one prefers to return to the deformed basis, it is straightforward to include the final application of $K_{\tilde{J}_z}$.

\section{Numerical examples}
\label{sec:results}

As a testing ground, we have chosen the phenomenological nuclear shell model, where a specified number of valence nucleons interact via one- and two-body interactions. Such a model has been used in recent applications to quantum computing by several groups \cite{Stetcu2022var,Kiss:2022,perezobiol2023nuclear,Li:2023}, and it has gained popularity because, while relatively simple to solve numerically with modest computational resources, it captures the essential complexity of nucleon-nucleon interactions. Of course, the motivation for quantum computing goes well beyond these simple examples to large-scale shell-model calculations that are not possible on classical hardware, and in our vision this is merely a testing ground.

Specifically, we will consider several examples in the $sd$ valence space consisting of the $1s_{1/2}0d_{3/2}0d_{5/2}$ single-particle harmonic oscillator states, which serves as a model for nuclei with up to 12 (valence) protons and 12 (valence) neutrons. The Hamiltonian we employ is the ``universal sd" effective interaction \cite{WILDENTHAL19845,*PhysRevC.74.034315}, which is fitted to reproduce the binding energies and excitation energies of nuclei in the $sd$ shell.

As in Refs. \cite{Stetcu2022var,Stetcu-2023proj}, we obtain the Hartree-Fock solution --- consisting of a single Slater determinant in a deformed basis --- using the code \textsc{sherpa} \cite{SHERPA}. The unitary transformation between the spherical and the deformed bases is then used to rewrite the interaction and other relevant operators ($J^2$, $J_x$, $J_y$, $J_z$) in the deformed basis [where we denote them by ($\tilde{J}^{2}$, $\tilde{J}_x$, $\tilde{J}_y$, $\tilde{J}_z$)]. The trial Hartree-Fock state can then be simply represented on a quantum computer: we adopt the standard Jordan-Wigner mapping \cite{JW}, so that the number of qubits is equal to the number of single-particle states. For each particle, one fills the lowest-energy available single-particle state, setting the corresponding qubit to state $|1\rangle$ and leaving the qubits for the unoccupied states in $|0\rangle$. Note that since we have two types of particles, the number of single-particle states in the problem is equal to the sum of number of states for each species (in the $sd$ shell, we have 12 single-particle states for protons and neutrons, respectively).

\subsection{Convergence properties}
\label{sec:convergence}

\begin{figure}[t]
\centering
\includegraphics[width=\columnwidth]{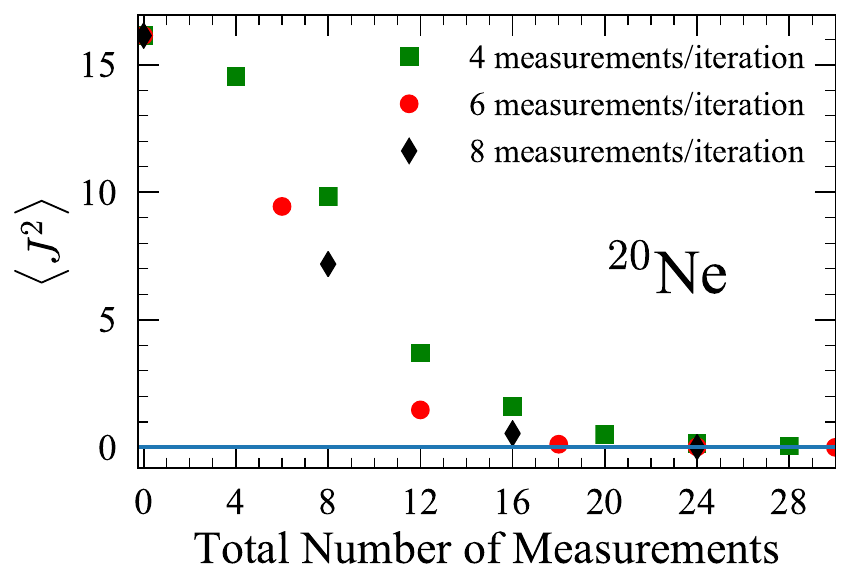}
\caption{\label{fig:ne20}
Expectation value of the total spin-squared operator $J^2$ as a function of the number of measurements performed. Each iteration requires four (green squares), six (red circles), or eight (black diamonds) measurements, respectively. See text for details.}
\end{figure}

We demonstrate the convergence of the proposed algorithm using the example of $^{20}$Ne, a nucleus consisting of 2 protons and 2 neutrons in the $sd$ valence shell. The extremal $|M|=J$ value that can be present in this system is $|M|=8$, requiring 4 consecutive projections with $t_i=\pi/2$, $\pi/4$, $\pi/8$, and $\pi/16$ in order to remove all allowed values of $M\neq 0$. Performing these projections for both $J_z$ and $J_x$ would require $N_\mathrm{proj}=8$ measurements of the ancilla per each iteration. As discussed in Sec. \ref{sec:projection}, it is not necessary to remove all possible $M\neq 0$ at each step of the algorithm. Alternately, one could employ 2 different time values $t_i=\pi/2$, $\pi/4$ at each iteration, leading to $N_\mathrm{proj}=4$ measurements per iteration that remove all allowed values of $|M|$ except for 4 and 8; or 3 different time values $t_i=\pi/2$, $\pi/4$, $\pi/8$ at each iteration, leading to $N_\mathrm{proj}=6$ measurements per iteration that remove all allowed values of $|M|$ except 8.  These three approaches are compared numerically in Fig. \ref{fig:ne20}. The initial state has expectation value $\braket{J^2}\approx 16\hbar^2$, reflecting the complicated mixture of total angular momentum eigenfunctions present in the deformed trial wave function. We see that all 3 calculations converge to a $J=0$ state after 24 total measurements have been performed, regardless of how those measurements are distributed across iterations. Thus, as the number of required mid-circuit measurements appears to be constant for a fixed system, we will compare in the next section the gate depth of each formulation.

\begin{table}
    \centering
    \begin{tabular}{ccccccc}
    \hline \hline
       Nucleus &  ~~$N_p$~~ & ~~$N_n$~~ &  ~~~~$\tilde{J}^2$~~~~ & ~~$\tilde{J}_x$~~ & ~~$\tilde{J}_y$~~ & ~~$\tilde{J}_z$~~\\
     \hline
      $^{20}$Ne & 2 & 2 & 24,599 & 220 & 200 & 220 \\
      $^{22}$Ne & 2 & 4 & 28,523 & 246 & 242 & 246 \\
      $^{24}$Mg & 4 & 4 & 40,610 & 288 & 262 & 288 \\
      $^{26}$Mg & 4 & 6 & 45,051 & 288 & 264 & 288 \\        
    \hline \hline
    \end{tabular}
    \caption{The number of Pauli strings required to encode the deformed-basis operators $\tilde{J}^2$, $\tilde{J}_x$, $\tilde{J}_y$, and $\tilde{J}_z$ in the Jordan-Wigner mapping for selected Ne and Mg isotopes. $N_p$ and $N_n$ denote the number of active (valence) nucleons. Note that the precise number of Pauli strings depends on the particular choice of the deformed Hartree-Fock state, but we expect that the relative order of magnitude between one- and two-body operators will remain very similar.}
    \label{tab:num_pauli_strings_j}
\end{table}

\subsection{Resource estimate}
\label{sec:resources}
For comparison purposes, we begin by estimating the resource requirements of the algorithm proposed in Ref. \cite{Stetcu-2023proj}, which relies on Trotterization of the unitary operator $\exp(-it_i \tilde{J}^{2}\otimes Y_a)$ with $\tilde{J}^2$ the total angular momentum operator expressed in the deformed basis. As such, the precise resource requirements in terms of gate counts will depend on the details of the deformed state. We will attempt a rough estimate of the number of \textsc{cnot}s required for an actual implementation. We first show in Table \ref{tab:num_pauli_strings_j} the number of Pauli strings necessary to encode the deformed-basis operators $\tilde{J}^2$, $\tilde{J}_x$, $\tilde{J}_y$, and $\tilde{J}_z$ for select Ne and Mg isotopes in the $sd$ shell. The precise forms of these operators depend on the details of the transformation between the deformed and the spherical bases. Nonetheless, the following general trends should remain true as one varies the Hartee-Fock state: The total number of Pauli strings required to encode the two-body operator $\tilde{J}^2$ is considerably larger than the number needed for one-body operators. In addition, the Pauli strings required to represent the two-body operator tend to be significantly more complex in terms of the number $N_c$ of non-identity single-qubit Pauli operators ($X$, $Y$, or $Z$) contained within each string. 

\begin{figure*}[ht]
    \centering
    \includegraphics[width=0.75 \textwidth]{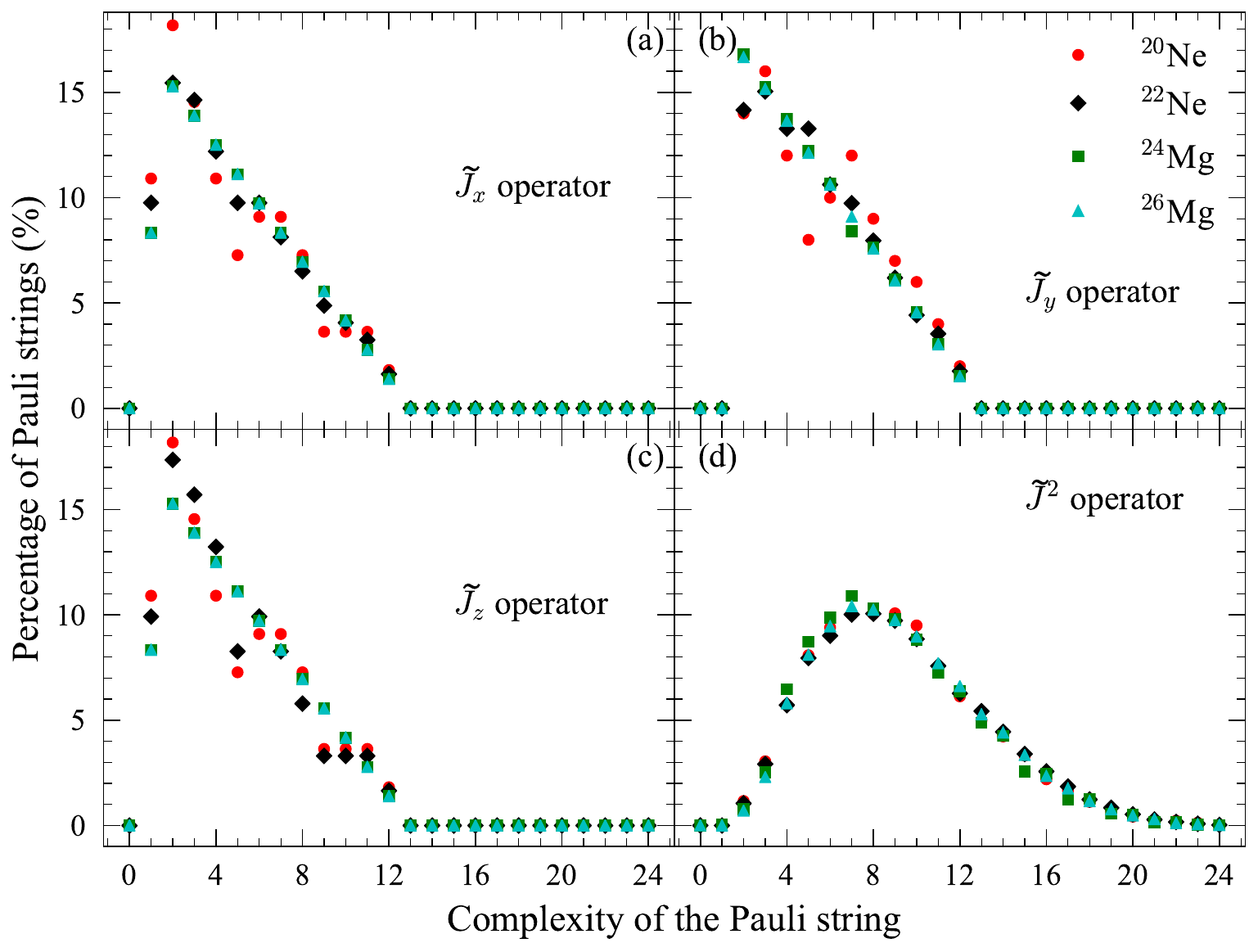}
    \caption{The percentage of Pauli strings with a given complexity necessary to encode the deformed-basis operators (a) $\tilde{J}_x$, (b) $\tilde{J}_y$, (c) $\tilde{J}_z$, and (d) $\tilde{J}^2$ for the Ne and Mg isotopes in Table \ref{tab:num_pauli_strings_j}. The complexity of a Pauli string is defined as the total number of non-identity Pauli operators ($X$, $Y$, and $Z$) appearing in the string. The maximum complexity is given by the number of qubits in the problem (in this case, 24 qubits representing 24 single-particle states).}
    \label{fig:complexityPauli}
\end{figure*}

Figure \ref{fig:complexityPauli} shows the relative percentage of Pauli strings of a given complexity resulting from Jordan-Wigner encodings of $\tilde{J}_x$, $\tilde{J}_y$, $\tilde{J}_z$, and $\tilde{J}^2$. For one-body operators in the $sd$ shell-model space, individual Pauli strings can contain at most 12 single-qubit Pauli $X$, $Y$ or $Z$ operators, a consequence of the factorization between protons and neutrons that these operators preserve. In contrast, the proton-neutron cross terms in $\tilde{J}^2$ generate a significant number of Pauli strings with more than 12 non-identity single-qubit operators. As the deformed-basis operators $(\tilde{J}_x,\tilde{J}_y,\tilde{J}_z)$ bear little resemblance to their spherical-basis counterparts $(J_x,J_y,J_z)$, there is no significant difference among the distribution of Pauli string complexities of $\tilde{J}_x$, $\tilde{J}_y$, and $\tilde{J}_z$. (In the spherical basis, the Pauli-string distributions would show a marked contrast. For example, unlike $J_x$ and $J_y$, the spherical-basis operator $J_z$ would contain only Pauli strings of complexity zero and one.) Across the four different $sd$-shell nuclei considered here, the general trends are very similar though there is some variation with isotope among the one-body operators. 

Knowing the complexity $N_c$ of the Pauli strings that encode $\tilde{J}^2$, we can estimate the number of \textsc{cnot} gates required to encode the unitary transformation $\exp(-it_i \tilde{J}^{2}\otimes Y_a)$. A ladder circuit that implements time evolution by a single Pauli string of complexity $N_c$ requires $2(N_c-1)$ \textsc{cnot}s \cite{doi:10.1080/00268976.2011.552441} (and additional single-qubit gates that we neglect here). As these are rough estimates, we do not attempt to simplify a circuit by cancelling consecutive \textsc{cnot}s. However, one does not expect that the cancellation will significantly reduce the number of \textsc{cnot}s.

When using the operator $\tilde{J}^2$, we remove $J>0$ states by performing measurements with evolution times $t_i=\pi/2^{i+1}$, thus eliminating states that satisfy $J(J+1)=(2m+1)2^i$ for some non-negative integer $m$ \cite{Stetcu-2023proj}. Therefore the longest propagation time required, which should govern the error incurred due to Trotterization, is $t_1=\pi/4$. As our primary focus is on the simplified algorithm introduced in this work, we do not attempt to quantify the number of Trotter steps required to adequately control the error associated with implementing the $\tilde{J}^2$ algorithm. Instead, we will limit ourselves to estimating the number of \textsc{cnot}s required to perform a single Trotter step of one projection evolution $\exp(-it_i \tilde{J}^{2}\otimes Y_a)$ for some fixed $t_i$, regarding this as the absolute minimal requirement to implement the $\tilde{J}^2$ algorithm, though the associated Trotter errors should be significant. As we will see, even this extreme minimal value of \textsc{cnot}s far exceeds the number required by our improved algorithm, which performs exact time evolution without the need for Trotterization. The cost of performing additional Trotter steps or further eliminations of remaining $J\neq 0$ states is almost certain to far outweigh any resources potentially saved from circuit optimization.

Given these caveats, Table \ref{tab:num_cnots} reports the estimated number of \textsc{cnot}s required to perform a single Trotter step of the $\tilde{J}^2$-based algorithm introduced in Ref. \cite{Stetcu-2023proj} for 4 selected $sd$-shell nuclei. The number of required \textsc{cnot}s nearly doubles as one goes from $^{20}$Ne (2 valence protons and 2 valence neutrons) to $^{26}$Mg (4 valence protons and 6 valence neutrons). Because the distribution of Pauli-string complexities does not change significantly between these four nuclei (see Fig. \ref{fig:complexityPauli}), the relative changes in the number of required \textsc{cnot}s are commensurate with the corresponding increases in total Pauli strings required to encode $\tilde{J}^2$ for each nucleus (see Table \ref{tab:num_pauli_strings_j}).

 \begin{table}[ht]
    \centering
    \begin{tabular}{ccccc}
    \hline \hline
      Nucleus  &  ~~$N_\mathrm{proj}$~~ & ~~$N_\mathrm{iter}$~~  & ~~$J_x$ \& $J_z$~~ & ~~$J^2$\\
      \hline
      $^{20}$Ne &4 & 10 & 3,944 & 510,398\\ 
       & 6 & 6 & 3,048 & 510,398 \\
       & 8 & 5 & 3,064 & 510,398\\
    \hline
      $^{22}$Ne & 4 & 10 & 3,944 & 530,536 \\
                & 6 & 6 & 3,048 & 530,536 \\
                & 8 & 5 & 3,064 & 530,536 \\
    \hline
    
      $^{24}$Mg & 4 & 10 & 3,944 & 734,022 \\
                & 6 & 6  & 3,048 & 734,022 \\
                & 8 & 5  & 3,064 & 734,022 \\
 \hline
      $^{26}$Mg & 4 &  10 & 3,944 & 835,904 \\
                & 6 & 6  & 3,048 & 835,904 \\
                & 8 & 5  & 3,064 & 835,904 \\        
    \hline \hline
    \end{tabular}
    \caption{The estimated number of \textsc{cnot} gates required to implement the present algorithm (column four) as compared to a single Trotter step of the algorithm introduced in Ref. \cite{Stetcu-2023proj} (column five) for 4 different $sd$-shell nuclei. To estimate the ``minimal'' cost of the $J^2$ algorithm only a single measurement/projection is performed, while in the case of the current algorithm there is a choice to partition $N_\mathrm{proj}$ measurements across $N_\mathrm{iter}$ iterations as either $(N_\mathrm{proj},N_\mathrm{iter})=(4,10)$, $(6,6)$, or $(8,5)$.}
    \label{tab:num_cnots}
\end{table}

We now turn our attention to the resource requirements of the algorithm proposed in this work. The number of \textsc{cnot} gates required to implement this new algorithm is much simpler to calculate, as it depends only on the model space under consideration, not on the number of particles or the details of the deformed state. Let $N_s$ be the number of particle species and $n$ be the number of single-particle states for each species. There are three sources of \textsc{cnot} gates:

\begin{enumerate}
    \item In general, the first unitary transformation $K^{\dag}_{\tilde{J}_z}$ involves $N_Sn(n-1)/2$ two-qubit operations requiring $N_sn(n-1)$ \textsc{cnot} gates.

\item As discussed in Sec. \ref{sec:cartan}, we exploit the block-diagonal form of the spherical-basis operator $J_x$ in order to factorize the transformation $K_{J_x}$ with respect to single-particle $j$. Thus, instead of the na\"ive requirement of $N_sn(n-1)$ \textsc{cnot}s per application of $K_{J_x}$, we need only 
\begin{equation}
   N_{J_x}\equiv N_s\sum_{j\in P}2j(2j+1)
\end{equation}
\textsc{cnot}s to rotate from the usual spherical basis (where $J_z$ is diagonal) to the spherical basis where $J_x$ is diagonal. In the $sd$ valence space with both protons and neutrons, this corresponds to 88 \textsc{cnot}s per each application of $K_{J_x}$ compared to 264 \textsc{cnot}s if one did not exploit the block-diagonal form of $J_x$. This transformation is repeated twice ($K_{J_x}$ and $K^{\dag}_{J_x}$) per each iteration of the algorithm, making the reduction in circuit depth even more significant. As explained in Sec. \ref{sec:cartan}, this same fact cannot be used to simplify the transformation $K_{\tilde{J}_z}$, as $\tilde{J}_z$ is generally not block-diagonal in the deformed basis. 

\item Encoding the time-evolution operator $\exp\left(-it_i J_z\otimes Y_a\right)$ requires 2 \textsc{cnot}s per each (non-ancilla) qubit. This operator is applied $N_\mathrm{proj}$ times per each iteration.

\end{enumerate}

Thus, the total \textsc{cnot} cost to implement $N_\mathrm{iter}$ iterations consisting of $N_\mathrm{proj}$ projections per iteration is
\begin{equation}
 N_\textsc{cnot} = n(n-1) N_s+2nN_sN_\mathrm{iter}N_\mathrm{proj}+2N_\mathrm{iter}N_{J_x}.
   \label{eq:cnot_cost}
\end{equation}
This assumes that the transformation $K^{\dag}_{\tilde{J}_z}$ is applied once at the start of the algorithm. If one prefers to also apply $K_{\tilde{J}_z}$ at the end of the circuit to return to the deformed basis, the first term in Eq. \eqref{eq:cnot_cost} is multiplied by a factor of two. On the other hand, if one prepares the trial wave function directly in the spherical basis, then the first term of Eq. \eqref{eq:cnot_cost} can be omitted. These modifications represent a fixed cost with respect to the total number of measurements performed, so the resulting change to the circuit depth is typically not very significant.

When using the one-body projection algorithm proposed in this paper, we will make the conservative assumption that we need to perform $\approx 40$ measurements total in order to ensure that the $J>0$ states are completely filtered out. This is almost double the number of measurements necessary for convergence in Fig. \ref{fig:ne20}. 

Table \ref{tab:num_cnots} shows that the number of \textsc{cnot}s required to implement the new algorithm is reduced by more than 2 orders of magnitude compared to the bare minimum requirements of the existing projection algorithm. Moreover, the resource requirements of the new algorithm are fixed for all nuclei within a specific model space, whereas the cost of the $J^2$ algorithm nearly doubles from $^{20}$Ne to $^{26}$Mg. On the other hand, the remaining advantages of the $J^2$ algorithm are (1) it can be applied to filter any total angular momentum $J$ and (2) it requires far fewer mid-circuit measurements. 

It is straightforward to estimate the single-qubit-gate requirements of the simplified algorithm. As shown in Fig. \ref{fig:circuit_sketch} (c), each application of the two-qubit gate $\exp[i\theta(X_iY_{i+1}-Y_iX_{i+1})]$ requires 4 $R_x$'s, 2 $R_z$'s, 2 \textsc{cnot}s, 2 $H$'s, and 2 $R_z(\theta)$'s, where $R_x$ and $R_z$ represent the fixed-angle rotations $R_x(\pi/2)$ and $R_z(\pi/2)$, respectively, and here $\theta$ denotes a generic angle. Each application of the ancilla-qubit projection $\exp(-it_iJ_z\otimes Y_a)$ requires 2 $R_x$'s, 2 \textsc{cnot}s, and 1 $R_z(\theta)$ for each qubit in the system. Therefore, the total counts of required single-qubit gates are
\begin{equation}
    \begin{split}
        N_H&= N_{R_z}= n(n-1)N_s+2N_\mathrm{iter}N_{J_x}, \\
        N_{R_x}&=2n(n-1) N_s+2nN_sN_\mathrm{iter}N_\mathrm{proj}+4N_\mathrm{iter}N_{J_x},\\
        N_{R_z(\theta)}&=\frac{1}{2}N_{R_x}.
    \end{split}
\end{equation}
We can further decompose these operators into the universal gate set $\{H,S,\textsc{cnot},T\}$ using the fact that
\begin{equation}
    R_x\sim HSH,~~~~~R_z\sim S,
\end{equation}
where $\sim$ denotes equivalence up to a global phase. The single-qubit rotation $R_z(\theta)$ can approximated by a unitary operator $U$ expressible in terms of $\{H,S,T\}$ gates to accuracy $\epsilon \geq ||R_z(\theta)-U||$ \cite{Selinger:2012pqc,Ross:2014okw}. The required numbers of $T$ gates is $K+4\log_2(1/\epsilon)$, where $K\approx 10$.

In estimating the resource requirements of the new algorithm, one should also account for the computational cost of the classical optimizations required to determine the parameters of $K_{J_x}$ and $K_{\tilde{J_z}}$. As discussed in Appendix \ref{app:cartan}, the relevant cost function and its gradient can be evaluated with $\mathcal{O}(k^4)$ and $\mathcal{O}(k^6)$ time complexity, respectively, where $k$ is the number of qubits acted on by the transformation $K$. By exploiting the block-diagonal form of $J_x$ in the spherical basis, the resulting value of $k=2j+1$ for each transformation $K^{(j)}_{J_x}$ is likely much smaller than the value of $k=\sum_{j\in P}(2j+1)$ for the initial transformation $K_{\tilde{J}_z}$. Hence, it is the optimization associated with $K_{\tilde{J}_z}$ --- which transforms from the deformed to the spherical basis, thus acting on all qubits for a given particle species --- that is likely to be the computational limitation.  Of course, if one formulates the trial wave function directly in the spherical basis (where $J_z$ is diagonal), then one can avoid this potentially costly optimization.

\section{summary}
\label{sec:summary}
We have proposed a simplified version of the projection filtering algorithm described in Ref. \cite{Stetcu-2023proj} to perform angular momentum projections on $J=0$ states, preserving the original success probability. This can have immediate applications for state preparation starting from mean-field solutions for even-even nuclei. The proposed algorithm fully exploits the factorization between protons and neutrons and uses the simple algebra of the one-body operators $J_x$ and $J_z$ to represent the required unitary evolutions in terms of a very efficient quantum circuit without the need for Trotterization. Our rough estimates indicate that the proposed algorithm requires at least two orders of magnitude fewer \textsc{cnot} gates than the full fledged algorithm, which uses the two-body operator $J^2$ for projection. 

In practice, it is often straightforward to encode the initial trial wave function on a quantum computer if one works in the deformed basis. Our new algorithm has the additional advantage that, after projection to the $J=0$ component, the system is naturally transformed to the spherical basis, which can be convenient for further manipulations of the prepared state. This algorithm should be well suited for early applications in fault-tolerant quantum computing, as the resource requirements are much more modest than those typically needed by full state-preparation methods. In future work, we intend to further explore the hardware implementation of our algorithm, in particular to determine the error and noise mitigation that are necessary for a satisfactory preparation of the projected state. 

At present, we could not devise a similarly efficient projection algorithm for total angular momentum other than $J=0$ or $J=1/2$. In principle, one can use the orthogonality of the Wigner D-matrices in order to project the rotated state in Eq. \eqref{eq:transform} onto a specific $\ket{J,M}$ component, though implementing the necessary integration over Euler angles through a linear combination of unitaries \cite{Childs:2012} will potentially require a large number of ancilla qubits and complicated multi-qubit controlled transformations.

\section*{Acknowledgments}
The authors are grateful to Alessandro Baroni for several helpful discussions. This work was carried out under the auspices of the National Nuclear Security Administration of the U.S. Department of Energy at Los Alamos National Laboratory under Contract No. 89233218CNA000001. ER is supported by the National Science Foundation under cooperative agreement 2020275. IS and JC gratefully acknowledge support by the Los Alamos Information Science and Technology Institute Rapid Response program and by the Advanced Simulation and Computing (ASC) program. This work was partially funded by the U. S. Department of Energy, Office of Science, Advanced Scientific Computing Program Office under FWP ERKJ382.  JC also acknowledges the Quantum Science Center for partial support of his work on this project.

\appendix

\section{More details on Cartan decomposition}
\label{app:cartan}
Any Hermitian, one-body fermionic operator $H$ can be expressed in second quantization as
\begin{equation}
    \begin{split}
H=\sum_{i}\braket{i|H|i}\hat{a}^\dag_i\hat{a}_i+&\sum_{i<j}\bigg\{\mathrm{Re}\left[\braket{i|H|j}\right]\left(\hat{a}^{\dag}_i\hat{a}_j+\hat{a}^{\dag}_j\hat{a}_i\right)\\
&+i\mathrm{Im}\left[\braket{i|H|j}\right]\left(\hat{a}^{\dag}_i\hat{a}_j-\hat{a}^{\dag}_j\hat{a}_i\right)\bigg\},
\end{split}
\end{equation}
where $i,j$ label the available single-particle states, which we encode into qubits using the Jordan-Wigner mapping
\begin{equation}
    \begin{split}
        \hat{a}^{\dag}_j &= \frac{1}{2}\left(X_j-iY_j\right)\underset{i<j}{\otimes} Z_i,\\
        \hat{a}_j &= \frac{1}{2}\left(X_j+iY_j\right)\underset{i<j}{\otimes}Z_i.
    \end{split}
    \label{eq:jw_trans}
\end{equation}
The diagonal part of the fermionic operator measures the occupation of state $i$,
\begin{equation}
    \hat{a}_i^\dag \hat{a}_i = \frac{1}{2}\left(I_i-Z_i\right),
\end{equation}
and the off-diagonal pieces can be expressed as
\begin{equation}
\begin{split}
    \hat{a}_i^\dag \hat{a}_j + \hat{a}_j^\dag \hat{a}_i&= \frac{1}{2}\left(\widehat{X_iX_j}+\widehat{Y_iY_j}\right),\\
    \hat{a}_i^\dag \hat{a}_j - \hat{a}_j^\dag \hat{a}_i&= \frac{i}{2}\left(\widehat{X_iY_j}-\widehat{Y_iX_j}\right),
    \end{split}
    \label{eq:jw_off_diag}
\end{equation}
where 
\begin{equation}
    \widehat{A_iB_j}\equiv A_iZ_{i+1}Z_{i+2}...Z_{j-1}B_j.
\end{equation}

Therefore, if the matrix elements of $H$ in the chosen single-particle basis are purely real\footnote{In the spherical basis, matrix elements of $J_x$ and $J_z$ are purely real, whereas matrix elements of $J_y$ are purely imaginary.}, then $H$ can be expressed in terms of the basis
\begin{equation}
    \mathfrak{m}=\mathrm{span}\left\{Z_j,\widehat{X_iX_j},\widehat{Y_iY_j}|i,j=1,2,\ldots,n,i<j\right\},
    \label{eq:m_alg_def}
\end{equation}
with dimension $|\mathfrak{m}|=n^2$, where $n$ is the number of qubits (and the number of single-particle states). We will hereafter assume that $\mathrm{Im}[\braket{i|H|j}]=0$ for all $i,j$, so that $H\in\mathfrak{m}$. 

Exponentiating these operators $e^{i\mathfrak{m}t}$ will generate commutators $[\mathfrak{m},\mathfrak{m}]$ that are not elements of $\mathfrak{m}$. In fact, they are always elements of the algebra
\begin{equation}
 \mathfrak{k}=\mathrm{span}\left\{\widehat{X_iY_j},\widehat{Y_iX_j}|i,j=1,2,\ldots,n,i<j\right\},
 \label{eq:k_alg_def}
\end{equation}
with dimension $|\mathfrak{k}|=n(n-1)$. Adding these operators to $\mathfrak{m}$ produces a closed algebra. Thus, all operators appearing in $e^{i\mathfrak{m}t}$ must be elements of the total algebra $\mathfrak{g}=\mathfrak{k}\oplus\mathfrak{m}$ satisfying $[\mathfrak{m},\mathfrak{m}]\subseteq \mathfrak{k}$, $[\mathfrak{k},\mathfrak{k}]\subseteq \mathfrak{k}$ and $[\mathfrak{k},\mathfrak{m}]\subseteq \mathfrak{m}$. Finally, we can define a maximal Abelian algebra within $\mathfrak{m}$, which we denote by $\mathfrak{h}$. Clearly, in the case considered
\begin{equation}
    \mathfrak{h}=\mathrm{span}\left\{Z_i|i=1,2,\ldots,n\right\},
\end{equation}
with dimension $|\mathfrak{h}|=n$. 

The so-called ``KHK'' theorem \cite{KHANEJA200111,10.1063/1.2008210} states that for any $m\in \mathfrak{m}$, there exists a $K\in e^{i\mathfrak{k}}$ and an $h\in \mathfrak{h}$ such that
\begin{equation}
    m=KhK^{\dag}.
\end{equation}
Consequently, we can replace unitary time evolution by 
\begin{equation}
    e^{-iHt}=Ke^{-iht}K^{\dag},
    \label{eq:KHK_unitary}
\end{equation}
so that after applying the unitary transformation $K^{\dag}$, evolution by an arbitrary time $t$ can be performed without the need for Trotterization, as all elements of $\mathfrak{h}$ commute. Since $K\in e^{i\mathfrak{k}}$, we can na\"ively parameterize it as $K=\exp\left(i\sum_i \theta_i k_i\right)$ where the $k_i$ are elements of the Pauli-string basis for $\mathfrak{k}$. However, this form may be very expensive to implement on quantum hardware, as the elements $k_i\in \mathfrak{k}$ generally do not commute. Equivalently and more conveniently, we can adopt the factorized product form \cite{Kokcu:2021ctj} 
\begin{equation}
\begin{split}
    K(\vec{\theta})&=\prod_i e^{i\theta_ik_i}\\
    &=\prod_{i<j}\exp\left[i\left(\theta^{(x)}_{i,j}\widehat{X_iY_j}+\theta^{(y)}_{i,j}\widehat{Y_iX_j}\right)\right],
    \end{split}
    \label{eq:K_ansatz_general}
\end{equation}
which is clearly advantageous for quantum computation as it eliminates the need for any Trotterization of $K$. Note that $[\widehat{X_iY_j},\widehat{Y_kX_\ell}]=0$ for all $i<j$, $k<\ell$. 

Generally, $K(\vec{\theta})$ depends on $|\mathfrak{k}|=n(n-1)$ independent parameters. We can further reduce the number of required parameters by noting that Eq. \eqref{eq:jw_off_diag} implies that the coefficients of $\widehat{X_iX_j}$ and $\widehat{Y_iY_j}$ are always equal for the Hermitian, one-body fermionic operators that we consider. Therefore, the relevant propagator is invariant under the rotation
\begin{equation}
    e^{i\pi/4\sum_jZ_j}e^{-iHt}e^{-i\pi/4\sum_jZ_j},
\end{equation}
which maps $X_i\rightarrow -Y_i$, $Y_i\rightarrow X_i$, $Z_i\rightarrow Z_i$. The right-hand side of Eq. \ref{eq:KHK_unitary} must also be invariant under this symmetry. However, 
\begin{equation}
\begin{split}
    e^{i\pi/4\sum_jZ_j}e^{i\theta^{(x)}_{i,j} \widehat{X_iY_j}}e^{-i\pi/4\sum_j Z_j}&=e^{-i\theta^{(x)}_{i,j} \widehat{Y_iX_j}},\\
    e^{i\pi/4\sum_jZ_j}e^{i\theta^{(y)}_{i,j} \widehat{Y_iX_j}}e^{-i\pi/4\sum_j Z_j}&=e^{-i\theta^{(y)}_{i,j} \widehat{X_iY_j}},
    \end{split}
\end{equation}
and therefore the symmetry is preserved only if $\widehat{X_iY_j}$ and $\widehat{Y_iX_j}$ have equal and opposite coefficients, $\theta^{(x)}_{i,j}=-\theta^{(y)}_{i,j}\equiv \theta_{i,j}$. The resulting simplified ansatz for $K$ is
\begin{equation}
    K(\vec{\theta})=\prod_{i<j}\exp\left[i\theta_{i,j}\left(\widehat{X_iY_j}-\widehat{Y_iX_j}\right)\right],
    \label{eq:K_ansatz_reduced}
\end{equation}
which depends on $n(n-1)/2$ independent parameters.

The form of the transformation $K$ in Eq. \eqref{eq:K_ansatz_reduced} has the unfortunate quality that the exponentials contain Pauli strings of complexity $N_c$ for all $2\leq N_c \leq n$, requiring a total of $2n(n^2-1)/3$ \textsc{cnot}s to be implemented on quantum hardware. The Cartan decomposition $\mathfrak{g}=\mathfrak{k}\oplus\mathfrak{m}$ is identical to that of the nearest-neighbor transverse-field XY model. We can therefore apply the results of Appendix H in Ref. \cite{Kokcu:2021ctj} in order to replace the ansatz in Eq. \eqref{eq:K_ansatz_reduced} by the ansatz represented in Fig. \ref{fig:circuit_sketch} (b), which contains only 2-qubit gates, $\exp[i\theta_{i,\ell}(\widehat{X_iY_{i+1}}-\widehat{Y_iX_{i+1}})]=\exp[i\theta_{i,\ell}(X_iY_{i+1}-Y_iX_{i+1})]$, many of which can be performed in parallel. The improved ansatz can be written explicitly as
\begin{equation}
    K(\vec{\theta})=\prod_{\ell=1}^{n-1}\prod_{i=\ell}^1 e^{i\theta_{i,\ell}\left(X_iY_{i+1}-Y_iX_{i+1}\right)},
\end{equation}
which still depends on $n(n-1)/2$ independent parameters, but the number of \textsc{cnot}s required to implement the transformation on hardware is thus reduced to $n(n-1)$. 

When the operator $J_x$ is written in Jordan-Wigner encoding, the diagonal part vanishes and the only non-vanishing off-diagonal terms are of the form $X_iX_{i+1}+Y_iY_{i+1}$, reflecting the fact that $J_x=(J_++J_-)/2$ can change the magnetic quantum number by only one unit. Moreover, it satisfies the symmetry $i\rightarrow n - i$. For $K_{J_x}$, this additional symmetry implies that $\theta_{i,\ell}=\theta_{i,n-1+i-\ell}$. In other words, if the transformation is organized in the triangular form depicted in Fig. \ref{fig:circuit_sketch} (b), then the parameters are symmetric about the vertical center line. As a result, the number of independent parameters required to specify $K_{J_x}$ is reduced from $n(n-1)/2$ to $n^2/4$.

In all cases, the parameters $\vec{\theta}$ that specify the unitary transformation $K(\vec{\theta})$ are obtained by finding any local extremum of the cost function
\begin{equation}
    f(\vec{\theta})=\mathrm{Tr}\left[K^{\dag}(\vec{\theta})vK(\vec{\theta}) H\right],
    \label{eq:cost_func}
\end{equation}
where $v\in\mathfrak{h}$ is chosen such that $e^{ivt}$ is dense in $e^{i\mathfrak{h}}$. For our purposes, it is sufficient to choose $v = \sum_i \gamma_i Z_i$, where the $\gamma_i$ are mutually irrational \cite{10.1063/1.2008210}. This optimization is performed classically: In general, elements of $\mathfrak{g}$ can be represented by matrices of size $2^n\times 2^n$. However, both the cost function $f(\vec{\theta})$ and its gradient $\partial f(\vec{\theta})/\partial\theta_i$ can be evaluated algebraically, requiring $\mathcal{O}(|\mathfrak{k}||\mathfrak{m}|)=\mathcal{O}(n^4)$ and $\mathcal{O}(|\mathfrak{k}|^2|\mathfrak{m}|)=\mathcal{O}(n^6)$ time complexity, respectively. See Appendix F of Ref. \cite{Kokcu:2021ctj} for details.

In principle, a Cartan decomposition of $J^2$ could be performed. However, the computational cost of the classical optimization would quickly become prohibitive as (1) the relevant algebra $\mathfrak{g}=\mathfrak{k}\oplus \mathfrak{m}$ is far more complex, so that the number of independent parameters required to specify $K$ grows exponentially with $n$ and (2) $J^2$ involves proton-neutron cross terms, limiting the extent that factorization can be exploited. 

By the same token, alternate encodings of the fermionic operators, such as Bravyi-Kitaev \cite{Bravyi:2000vfj}, will in general not map onto the simple algebras defined in Eqs. \eqref{eq:m_alg_def} and \eqref{eq:k_alg_def}. While one could still devise a projection algorithm using $J_x$ and $J_z$ in these encodings, one could not apply the same Cartan decomposition discussed here, which enables efficient application of the required rotations.

\section{Numerical values of transformation parameters}
\label{app:numerical_params}
As discussed above, if one works in the usual spherical basis, then the transformation $K_{J_x}$ that diagonalizes $J_x$ can be factorized with respect to single-particle angular momentum $j$. Moreover, the transformations $K_{J_x}^{(j)}$ are defined with respect to the spherical basis where $J_z$ is diagonal; therefore, they do not depend on the specifics of the deformed basis, and the parameters that specify them are generic. The KHK theorem holds at any local extremum of the cost function in Eq. \eqref{eq:cost_func}. As such, many different parameterizations are possible, and here we provide only one realization. 

We assume that the qubits labeled $i=1,2,\ldots,n$ correspond to single-particle states with increasing magnetic projection $m=\{-j,-j+1,\ldots,j-1,j\}$. A set of valid $\vec{\theta}$ specifying the transformations $K_{J_x}^{(j)}(\vec{\theta})$ as defined in Eq. \eqref{eq:K_op_basis} for $j=1/2$, $3/2$, $5/2$, and $7/2$ are given in Table \ref{tab:KJx_params}. They are presented in ``triangular form'', in analogy with the representation used in Fig. \ref{fig:circuit_sketch} (b). For example, in the case of $j=3/2$, the parameters $\theta_{i,\ell}$ that specify $K_{J_x}^{(3/2)}$ are
\begin{equation}
\begin{split}
    \theta_{1,1}=\theta_{1,3}&=1.8326,~~ \theta_{2,2}=\theta_{2,3}=1.1423,\\
    \theta_{1,2}&=1.3275,~~\theta_{3,3}=0.9661.
    \end{split}
    \label{eq:theta_params_j32}
\end{equation}

\begin{table*}[]
    \centering
    \begin{tabular}{c|ccc|ccccc|ccccccc}
    \hline
    \hline
          $j=1/2$     &   $j=3/2$      &        &        &    $j=5/2$     &         &         &         &         &   $j=7/2$      &         &         &         &         &         & 0.8297 \\
               &         &        &        &         &         &         &         &         &         &         &         &         &         &  0.9039 &        \\
               &         &        &        &         &         &         &         &  0.8743 &         &         &         &         &  1.0012 &         &  1.0317 \\
               &         &        &        &         &         &         & 0.9920  &         &         &         &         &  1.1038 &         &  1.1453 &        \\
               &         &        & 0.9661 &         &         &  1.1199 &         &  1.1500 &         &         &  1.2016 &         &  1.2380 &         &  1.2463 \\
               &         & 1.1423 &        &         &  1.2413 &         & 1.2711  &         &         &  1.2943 &         &  1.3198 &         &  1.3283 &        \\
      0.3927   &  1.8326 &        & 1.3275 &  1.7811 &         &  1.3788 &         &  1.7584 &  1.7515 &         &  1.4045 &         &  1.7311 &         &  1.4123 \\
      \hline
      \hline
    \end{tabular}
    \caption{Parameters $\theta_{i,\ell}$ specifying the transformation $K^{(j)}_{J_x}(\vec{\theta})$ for various single-particle spaces indexed by total angular momentum $j$. The parameters are organized to correspond to the triangular layout depicted in Fig. \ref{fig:circuit_sketch} (b). As discussed above, the parameters are symmetric about the center line, and here we present only the left half (plus center) of the triangle. [See the example in Eq. \eqref{eq:theta_params_j32} for more details.]}
    \label{tab:KJx_params}
\end{table*}

\bibliography{references}

\end{document}